\documentstyle[preprint,revtex]{aps}

\tightenlines

\begin{document}
\draft
\preprint{}
\begin{title}
Exact Universal Amplitude Ratios for Two-Dimensional
 \\Ising Models and a Quantum Spin Chain
\end{title}

\author{N. Sh. Izmailian  and   Chin-Kun Hu\cite{huck}}

\begin{instit}
Institute of Physics, Academia Sinica,
Nankang, Taipei 11529, Taiwan
\end{instit}

\begin{abstract}
Let $f_N$ and $\xi_N^{-1}$ represent, respectively, the free energy
per spin and the inverse spin-spin correlation length of the
critical Ising model on a $N \times \infty$ lattice,
with $f_N \to f_{\infty}$ as $N \to \infty$.
We obtain analytic expressions for $a_k$ and $b_k$ in the expansions:
 $N(f_N-f_{\infty})=\sum_{k=1}^{\infty}{a_k}/{N^{2 k-1}}$ and
$\xi_N^{-1}=\sum_{k=1}^{\infty}b_k/N^{2 k-1}$
for square, honeycomb, and plane-triangular lattices,
and find that
$b_k/a_k=(2^{2k}-1)/(2^{2k-1}-1)$ for all of these lattices,
 i.e. the amplitude ratio $b_k/a_k$ is universal.
We also obtain similar result for a critical quantum spin chain
and find that such result could be understood from a
perturbated conformal field theory.
\end{abstract}

\vskip 5 mm

\vskip 10 cm

\newpage

Experimental data, analytic and simulational studies of phase
transition models, and renormalization group (RG) theory suggest
that critical systems can be grouped into universality classes so
that the systems in the same class have the same set of critical exponents
\cite{stanley71,pf84,ahp91}.
RG theory has also been used to propose that
critical systems of the same universality class could have universal
finite-size scaling functions (UFSSF's) and universal amplitude ratios
\cite{pf84,ahp91,aharony} and some analytic and numerical calculations of
critical systems have supported the idea of universal
amplitude ratios \cite{pf84,ahp91,aharony}.
Using Monte Carlo methods \cite{hu92} and choosing appropriate
aspect ratios for lattices of critical systems, Hu, {\em et al.}
have found UFSSF's for percolation and Ising models \cite{huetal}.
Despite the success of RG theory and Monte Carlo simulations,
it is valuable to have more analytic results which could widen or deepen
our understanding of the universality of critical systems.
In this Letter we present exact calculations for a set of
universal amplitude ratios for the Ising model on
square (sq), honeycomb (hc), and plane-triangular (pt) lattices
\cite{onsager,wannier,husimi} and for a quantum spin chain
\cite{katsura62}, which is in
the universality class of two-dimensional (2D) Ising models.
As far as we know,
no previous RG arguments, analytic calculations, or numerical studies
predict the existence of this whole set of universal amplitude ratios.

Let $f_N$ and $\xi_N^{-1}$ represent, respectively, the free energy
per spin and the inverse spin-spin correlation length of the Ising model
\cite{onsager,wannier,husimi} on a $N \times \infty$ lattice with
periodic boundary conditions,
with  $f_N \to f_{\infty}$ as $N \to \infty$.
In this letter,
we obtain analytic equations for $a_k$ and $b_k$ in the expansions:
\begin{equation}
N(f_N-f_{\infty})=\sum_{k=1}^{\infty}\frac{a_k}{N^{2 k-1}},
\label{fN}
\end{equation}

\begin{equation}
\xi_N^{-1}=\sum_{k=1}^{\infty}\frac{b_k}{N^{2 k-1}},
\label{cli}
\end{equation}
for sq, hc, and pt lattices, and find that
\begin{equation}
b_k/a_k=(2^{2k}-1)/(2^{2k-1}-1),
\label{I4}
\end{equation}
for all of these lattices, i.e.
the amplitude ratio $a_k/b_k$ is universal.
We also obtain similar expansions for the critical
ground state energy $E_0$
and the critical first energy gap ($E_1-E_0$)
of a quantum spin chain \cite{katsura62},
which are, respectively, the quantum analogies of the free
energy and inverse spin-spin correlation length for the Ising model,
and find that the amplitude ratios have the same values.
We could physically understand  such result from a
perturbated conformal field theory.

Consider an Ising ferromagnet on an $N \times M$ lattice
with periodic boundary conditions ({\it i.e.} a torus).
The Hamiltonian of the system is
\begin{equation}
\beta H=-J\sum_{<ij>} s_i s_j,
\label{I5}
\end{equation}
where $\beta = (k_BT)^{-1}$, the Ising spins $s_i=\pm 1$ are located at
the sites of the lattice and the summation goes over all
nearest-neighbor pairs of the lattice.
We consider a transfer matrix acting along the $M$ direction
\cite{domb,night76,derrida}.
If $\Lambda_0$ and $\Lambda_1$ are the largest and the second-largest
 eigenvalues of the transfer matrix, in the limit $M \to \infty$
the free energy per spin, $f_N$, and the inverse
longitudinal spin-spin correlation length, $\xi_N^{-1}$,
are

\begin{equation}
f_N=\frac{1}{\zeta N} \ln{\Lambda_0}
\quad \mbox{and} \quad
\xi_N^{-1}=\frac{1}{\zeta}\ln{(\Lambda_0/\Lambda_1)}.
\label{I6}
\end{equation}
Here $\zeta$ is a geometric factor which is $1$, $2/\sqrt{3}$ and
$1/\sqrt{3}$ for sq, hc, and pt lattices,
respectively  \cite{pf84}.
Exact expressions for eigenvalues
$\Lambda_0$ and $\Lambda_1$ are available for all lattices under consideration: sq
\cite{onsager,domb,night76,derrida}, hc \cite{pf84,husimi},
and pt \cite{wannier}.

We start from the Ising model on the sq lattice. Onsager
\cite{onsager} has obtained expressions for all eigenvalues of the
transfer matrix. The two leading eigenvalues are
$\Lambda_0 =  (2 \sinh 2 J)^{N/2}
\exp{\left(\frac{1}{2}\sum_{r=0}^{N-1} \gamma_{2 r +1}\right)}$,
$\Lambda_1 = (2 \sinh 2 J)^{N/2}
\exp{\left(\frac{1}{2}\sum_{r=1}^N \gamma_{2 r}\right)}$,
where $\gamma_k$  is implicitly given by
$\cosh{\gamma_k}=\cosh{2 J}\coth{2 J}-\cos{(k \pi/N)}$.
At the critical point $J_c$ of the sq lattice Ising model,
where $J_c=\frac{1}{2}\ln{(1+\sqrt{2})}$,
one then obtains  $\gamma_k= 2 \psi_{sq} \left(\frac{k \pi}{2 N}\right)$.
Here
\begin{equation}
\psi_{sq}(x)=\ln{\left(\sin{x}+\sqrt{1+\sin^2{x}}\right)}.
\label{I7}
\end{equation}
Then the critical free energy $f_ N$ and critical spin-spin correlation length
$\xi_N$ of Eq. (\ref{I6}) can be written as
\begin{eqnarray}
f_N&=& \frac{1}{2} \ln{2}+\frac{1}{2 N} \sum_{r=0}^{N-1} \gamma_{2 r+1},
\label{I81} \\
\xi_N^{-1}&=&\frac{1}{2}\sum_{r=0}^{N-1}(\gamma_{2 r+1}-\gamma_{2 r}).
\label{I8}
\end{eqnarray}
It is readily seen from Eqs. (\ref{I7}), (\ref{I81})
 and (\ref{I8}) that $\xi_N^{-1}$ and $N f_N$ have odd
parity as a function of $N^{-1}$. Therefore, in the
following expansions of $\xi_N^{-1}$ and $N f_N$
as a function of $N^{-1}$, we keep only odd terms.

To write $f_N$ and $\xi_N^{-1}$ in the form of Eqs. (\ref{fN}) and
(\ref{cli}),
we must evaluate Eqs. (\ref{I81}) and (\ref{I8}) asymptotically. These sums
can be handled by using the Euler-Maclaurin summation formula \cite{hardy}.
After a straightforward calculation, we have
\begin{eqnarray}
N (f_N-f_{\infty})&=&
\sum_{k=1}^{\infty}\frac{2 B_{2 k}}{(2 k)!}(2^{2 k-1}-1)
\left(\frac{\pi}{2 N}\right)^{2 k-1}\psi_{sq}^{(2 k-1)}
\nonumber \\
&=&
\frac{\pi}{12 N}+\frac{7}{180}\left(\frac{\pi}{2 N}\right)^3
 +\frac{31}{756} \left(\frac{\pi}{2 N}\right)^5
+\frac{10033}{75600}\left(\frac{\pi}{2 N}\right)^7
+\dots,
\label{as} \\
\xi_N^{-1}&=&\sum_{k=1}^{\infty}\frac{2 B_{2 k}}{(2 k)!}(2^{2 k}-1)
\left(\frac{\pi}{2 N}\right)^{2 k-1}
\psi_{sq}^{(2 k-1)}
\nonumber\\
&=&
\frac{\pi}{4 N} +\frac{1}{12}\left(\frac{\pi}{2 N}\right)^3
 +\frac{1}{12} \left(\frac{\pi}{2 N}\right)^5
+\frac{1343}{5040}\left(\frac{\pi}{2 N}\right)^7 +\dots.
\label{bs}
\end{eqnarray}
Here $B_{2 k}$ are the Bernoulli
numbers and
$\psi_{sq}^{(2 k-1)} = \left(d^{2 k-1}
\psi_{sq}(x)/dx^{2 k-1}\right)_{x=0}$; $b_2 = \pi^3/96$
has been computed previously by Derrida and Seze \cite{derrida}.

For the Ising model on the honeycomb (hc) lattice, 
Husimi and Syozi \cite{husimi} found that 
$\Lambda_0=(2 \sinh 2J)^N \exp{(\gamma_1+\gamma_3+...+\gamma_{N-1})}$ and
$\Lambda_1=(2 \sinh 2 J)^N \exp{(\frac{1}{2}\gamma_0+\gamma_2+
...+\gamma_{N-2}+\frac{1}{2}\gamma_{N})}$, where the $\gamma_r$ are given by
$\cosh{\gamma_r} = \cosh{2 J}\cosh{2 J^*}-\sin^2{\frac{\pi r}{N}}
-\cos{\frac{\pi r}{N}}(\sinh^2{2 J} \sinh^2{2 J^*}
 -\sin^2{\frac{\pi r}{N}})^{1/2}$. Here $J^*$ is defined by
$(\cosh{2 J}-1)(\cosh{2 J^*}-1)=1$, so that one has $J^*=J$ at the critical
point ($J_c=\frac{1}{2}\ln{(2+\sqrt{3})}$) and one then obtains
$\gamma_r = 2 \psi_{hc}\left(\frac{r \pi}{2 N}\right)$, where

\begin{equation}
\psi_{hc}(x)=\ln{\left\{A(x)+\sqrt{A^2(x)-1}\right\}},
\label{psihc}
\end{equation}
with $A(x)=\left(\sqrt{8+\cos^2{2 x}}-\cos{2 x}\right)/2$.
Using the Euler-Maclaurin summation formula,
we can write the free energy $f_N$ and the inverse spin-spin correlation
length $\xi_N^{-1}$ for the hc lattice as
\begin{eqnarray}
N (f_N &-& f_{\infty}) =
\sum_{k=1}^{\infty}\frac{\sqrt{3} B_{2 k}(2^{2 k-1}-1)}{(2 k)!}
\left(\frac{\pi}{2 N}\right)^{2 k-1}\psi_{hc}^{(2 k-1)}
\nonumber \\
&=&\frac{\pi}{12 N}
 -\frac{31}{210} \left(\frac{\pi}{3 N}\right)^5
+\frac{511}{110}\left(\frac{\pi}{3 N}\right)^9+...,
\label{I15} \\
\xi_N^{-1}&=&
\sum_{k=1}^{\infty}\frac{\sqrt{3} B_{2 k}(2^{2 k}-1)}{(2 k)!}
\left(\frac{\pi}{2 N}\right)^{2 k-1}
\psi_{hc}^{(2 k-1)}
\nonumber\\
&=& \frac{\pi}{4 N}
 -\frac{3}{10} \left(\frac{\pi}{3 N}\right)^5
+\frac{93}{10}\left(\frac{\pi}{3 N}\right)^9+...,
\label{I16}
\end{eqnarray}
where  $\psi_{hc}^{(2 k-1)} = \left(d^{2 k-1}
\psi_{hc}(x)/dx^{2 k-1}\right)_{x=0}$.

For the case of the pt lattice, we note that one can use the
star-triangle transformation to transform the hc to the pt
lattice \cite{domb}. The amplitudes of the $N^{-3}$ and $N^{-7}$
correction terms are identically zero for the hc and the pt lattices.

Above results for the sq, hc, and pt lattices
can be summarized as
\begin{equation}
N (f_N - f_{\infty})=
\sum_{k=1}^{\infty}\frac{2 B_{2 k}(2^{2 k-1}-1)}{\zeta (2 k)!}
\left(\frac{\pi}{2 a N}\right)^{2 k-1}\psi^{(2 k-1)},
\label{fN-all}
\end{equation}

\begin{equation}
\xi_N^{-1} =
\sum_{k=1}^{\infty}\frac{2 B_{2 k}(2^{2 k}-1)}{\zeta (2 k)!}
\left(\frac{\pi}{2 a N}\right)^{2 k-1}
\psi^{(2 k-1)},
\label{I18}
\end{equation}
where $a = 1, \quad \zeta = 1, \quad \psi(x) = \psi_{sq}(x)$
for the sq lattice,
 $a = 1, \quad \zeta = 2/\sqrt{3}, \quad \psi(x) = \psi_{hc}(x)$ for
the hc lattice, and $a = 2, \quad \zeta = 1/\sqrt{3}, \quad \psi(x) =
\psi_{tr}(x) = \psi_{hc}(x)$ for the pt lattice.
The ratios of the nonvanishing amplitudes of the $N^{-(2k-1)}$
correction terms in the spin-spin correlation length and the free energy
expansion, {\it i.e.} $b_k/a_k$,
are the same for all three lattices under consideration \cite{anisotropy}.
Thus we have established Eq. (\ref{I4}).

To check whether Eq. (\ref{I4}) is still valid for other models
in the Ising universality class, we proceed
to study a quantum spin model on a one-dimensional lattice of $N$ sites
with periodic boundary conditions, whose Hamiltonian is \cite{katsura62}

\begin{equation}
H=-\frac{\lambda}{2 \gamma}\sum_{n=1}^N \sigma^z_n
-\frac{1}{4 \gamma}\sum_{n=1}^N\left[(1+\gamma)\sigma^x_{n+1}\sigma^x_n+
(1-\gamma)\sigma^y_{n+1}\sigma^y_n\right],
\label{chain}
\end{equation}
where $\sigma^x, \sigma^y$ and $\sigma^z$ are the Pauli matrices. The phase
diagram is well known \cite{bm71}. For all $\gamma$ $(0<\gamma \le 1)$,
there is a critical point at $\lambda_c =1$, which falls into the
two-dimensional Ising universality class. The inverse correlation length
$\xi_i^{-1}$ is given by the difference in eigenvalues $E_i-E_0$ of the
Hamiltonian $H$. In particular, the first energy gap gives the
inverse spin-spin correlation length $\xi_1^{-1}$
and the second energy gap is the inverse energy-energy correlation length
$\xi_2^{-1}$.
By expanding the exact solution of Eq. (\ref{chain}),
Henkel \cite{henkel87} has
obtained several finite-size correction terms to the ground-state
 energy $E_0$, to the first ($E_1-E_0$) and second ($E_2-E_0$) energy gaps.
We have extended the calculations to arbitrary order and found that
\begin{eqnarray}
-E_0- N \alpha_0 &=&
\sum_{k=1}^{\infty}\frac{2 B_{2 k}(2^{2 k-1}-1)}{(2 k)!}
\left(\frac{\pi}{2 N}\right)^{2 k-1}\psi_{q}^{(2 k-1)}
\label{E0} \\
&=&\frac{\pi}{12 N}
-\frac{7}{15}\left(\frac{1}{\gamma^2}-\frac{4}{3}\right)
\left(\frac{\pi}{4 N}\right)^3-\frac{62}{63}
\left(\frac{1}{\gamma^4}-\frac{16}{15}\right) \left(\frac{\pi}{4 N}\right)^5
+\dots,
\nonumber\\
E_1-E_0&=&
\sum_{k=1}^{\infty}\frac{2 B_{2 k}(2^{2 k}-1)}{(2 k)!}
\left(\frac{\pi}{2 N}\right)^{2 k-1}
\psi_{q}^{(2 k-1)}
\nonumber\\
&=&\frac{\pi}{4 N}-\left(\frac{1}{\gamma^2}-\frac{4}{3}\right)
\left(\frac{\pi}{4 N}\right)^3
-2 \left(\frac{1}{\gamma^4}-\frac{16}{15}\right)
\left(\frac{\pi}{4 N}\right)^5 +\dots,
\label{E10}\\
E_2-E_0&=&
\sum_{k=1}^{\infty}\frac{8 k}{(2 k)!}
\left(\frac{\pi}{2 N}\right)^{2 k-1}
\psi_{q}^{(2 k-1)}
\nonumber\\
&=&\frac{2 \pi}{N}+16\left(\frac{1}{\gamma^2}-\frac{4}{3}\right)
\left(\frac{\pi}{4 N}\right)^3
-16 \left(\frac{1}{\gamma^4}-\frac{16}{15}\right)
\left(\frac{\pi}{4 N}\right)^5 +\dots,
\label{enerener2}
\end{eqnarray}
where ,  $\psi_{q}^{(2 k-1)} =
\left(d^{2 k-1} \psi_{q}(x)/dx^{2 k-1}\right)_{x=0}$,
$\psi_{q}(x)=\sqrt{\sin^2{x}-(1-1/\gamma^2)\sin^4{x}}$,
and $\alpha_0$ is an non-universal number
$\alpha_0 = \frac{2}{\pi}\int_0^{\pi} \psi_{q}(x) d x
= 2 \left[1+\arccos{\gamma}/(\gamma
\sqrt{1-\gamma^2})\right]/\pi$.
Thus, the ratios of amplitudes for ($E_1-E_0$) and ($- E_0$)
also satisfy Eq. (\ref{I4}). Equations (\ref{E0}) and
(\ref{enerener2}) implies also
that the ratios $\bar r_k$ of amplitudes for
($E_2-E_0$) and ($- E_0$) are $\gamma$-independent and given by
\begin{equation}
\bar r_k =\frac{4 k}{(2^{2 k-1}-1)B_{2 k}}.
\label{enerener}
\end{equation}

It is of interest to compare this finding with other results.
The exact and numerical estimates \cite{queiroz} of the subdominant
correction amplitudes for the sq, hc and pt lattices
are presented in Table I, which shows that the numerical values obtained
by Queiroz \cite{queiroz} are very close to our exact results.
On the basis of conformal invariance,
the asymptotic finite-size
scaling behavior of the critical free energy and the inverse correlation
length is found to be \cite{affleck}
\begin{equation}
\lim_{N \to \infty} {N^2 (f_N-f_{\infty})}=\frac{c \pi}{6},
\label{I2}
\end{equation}
\begin{equation}
\lim_{N \to \infty} {N \xi_i^{-1}}=\lim_{N \to \infty} {N (E_i-E_0)} =
2 \pi x_i,
\label{I1}
\end{equation}
where $c$ is the conformal anomaly number and $x_i$ is the
scaling dimension of the $i$th scaling field. For the 2D Ising model,
we have $c=1/2$, $x_1=\eta/2=1/8$ and $x_2=1$ and the leading terms
of Eqs. (\ref{fN-all}) and (\ref{I18})
for all of the sq, hc, and pt lattices are consistent with
Eqs. (\ref{I2}) and (\ref{I1}),
{\it i.e.} $a_1$ and $b_1$ are universal.
Equations (\ref{I2}) and (\ref{I1}) implies
immediately that their ratio is also universal, namely
\begin{equation}
\lim_{N \to \infty} \frac{E_1-E_0}{N (f_N-f_{\infty})}=r_1 \qquad \mbox{and}
\qquad \lim_{N \to \infty} \frac{E_2-E_0}{N (f_N-f_{\infty})}=\bar r_1,
\label{I3}
\end{equation}
where $r_1=12 x_1/c$ and $\bar r_1=12 x_2/c$. For the 2D Ising universality
class we have
$r_1=3, \, \bar r_1=24$, which is consistent with Eqs. (\ref{I4}) and
(\ref{enerener})
for the case $k=1$.

The corrections to Eqs. (\ref{I2}) and (\ref{I1})  can be calculated by
the means of a perturbated conformal field theory \cite{cardy86,zamol87}.
In general, any lattice Hamiltonian will contain correction terms to the
critical Hamiltonian $H_c$
\begin{equation}
H = H_c + \sum_p g_p \int_{-N/2}^{N/2}\phi_p(v) d v,
\label{Hc}
\end{equation}
where $g_p$ is a non-universal
constant and $\phi_p(v)$ is a perturbative conformal field.
Below we will consider the case with only one perturbative conformal field,
say $\phi_l(v)$.
Then the eigenvalues of $H$ are
\begin{equation}
E_n=E_{n,c}+g_l \int_{-N/2}^{N/2}<n|\phi_l(v)|n> d v + \dots,
\label{En}
\end{equation}
where $E_{n,c}$ are the critical eigenvalues of $H$. The matrix element
$<n|\phi_l(v)|n>$ can be computed in terms of the universal structure
constants $(C_{nln})$ of the operator product expansion \cite{cardy86}:
$<n|\phi_l(v)|n> =\left({2 \pi}/{N}\right)^{x_l}C_{iln}$,
where $x_l$ is the scaling dimension of the conformal field $\phi_l(v)$.
The correlation lengths $(\xi_n^{-1}=
E_n-E_0)$ and the ground-state energy ($E_0$) can be written as
\begin{eqnarray}
\xi_n^{-1}&=&\frac{2 \pi}{N} x_n+
2 \pi g_l(C_{nln}-C_{0l0})\left(\frac{2 \pi}{N}\right)^{x_l-1} + \dots,
\label{xin}\\
E_0 &=& E_{0,c}+2 \pi g_l C_{0l0}
\left(\frac{2 \pi}{N}\right)^{x_l-1} + \dots.
\label{E0conf}
\end{eqnarray}
Equations (\ref{xin}) and (\ref{E0conf}) show that while the amplitude
of correction to scaling terms are not universal, ratios of them are.
For the 2D Ising model, one finds \cite{henkel}
that the leading finite-size corrections ($1/N^3$) can be described by the
Hamiltonian given by Eq. (\ref{Hc}) with a single
perturbative conformal field $\phi_l(v)=L_{-2}^2(v)+{\bar L}_{-2}^2(v)$
with scaling dimension $x_l=4$ .
The universal structure constants $C_{2l2}$, $C_{1l1}$ and $C_{0l0}$ can be
obtained from the matrix element
$<n|L_{-2}^2(v)+{\bar L}_{-2}^2(v)|n>$, which have already been computed
by Reinicke \cite{reinicke87} ($C_{2l2}=1729/5760$, $C_{1l1}=-7/720$ and
$C_{0l0}=49/5760$). Equations (\ref{xin}) and (\ref{E0conf}) implies that
the ratios of first-order corrections amplitudes for ($E_n-E_0$) and
($- E_0$) is universal and equal to $(C_{0l0}-C_{nln})/C_{0l0}$,
which is consistent with Eq. (\ref{I4}) and Eq. (\ref{enerener}) for the
cases $n=1, k=2$ $[(C_{0l0}-C_{1l1})/C_{0l0} = 15/7]$ and
$n=2, k=2$ $[(C_{0l0}-C_{2l2})/C_{0l0} = -240/7]$ respectively.
Compare the  amplitudes of the $N^{-3}$ correction terms
for the Ising model and the quantum spin chain with the general results of
Eqs. (\ref{xin}) and (\ref{E0conf}) one can find that
$g_l=(3/{\gamma}^2 - 4)/56 \pi$ for the quantum spin chain and
$g_l=-1/28 \pi$ for the Ising model on the sq lattice. For the Ising model
on the hc and pt lattices we find that $g_l=0$, which indicate that at least
two perturbative conformal fields are necessary to generate
all finite-size correction
terms. Further work has to be done to possibly evaluate exactly all
finite-size correction terms from perturbative conformal field theory.

The results of this Letter inspire several problems for further
studies: (i) On the basis of perturbated conformal field theory, can one
find other universal amplitude ratios? (ii) How do such amplitudes
behave in other
models, for example in the three-state Potts model?
(iii) For the critical Ising model on a large $N \times M$ sq lattice
($M/N$ is a finite number), we have obtained expansions in $N^{-1}$
for the free energy, the internal energy, and the specific
heat \cite{hu}. It is of interest to extend such expansions
to inverse spin-spin correlation length and to hc and pt lattices, and
to study whether the amplitude ratios are also universal.

We would like to thank I. Affleck, M. Henkel, E. V. Ivashkevich and
S. L. A. de Queiroz for valuable comments and Jonathan Dushoff
for a critical reading of the paper.
This work was supported in part by the National Science Council of the
Republic of China (Taiwan) under Contract No. NSC 89-2112-M-001-005.

\begin{table}
\caption{Comparison of exact (Eqs. (\ref{as}), (\ref{bs}),
(\ref{I15}) - (\ref{I18})) and numerical \cite{queiroz} values for
subdominant finite-size corrections terms in free energy and
inverse spin-spin correlation length expansion.}
\end{table}


\begin{references}
\bibitem{anisotropy} To check the applicability of these surprising results,
we study the anisotropic sq lattice Ising model with coupling constant $J$ and $kJ$
along the horizontal and vertical directions, respectively, with $0 < k < \infty$.
At the critical point $J_c$, where $J_c$ is defined 
by $\sinh{2J_c}\sinh{2kJ_c} = 1$, we obtain
$\psi_{sq}(x)=\ln{\left[\sin{x}/\sinh{2J_c}
+\sqrt{1+(\sin{x}/\sinh{2J_c})^2}\right]}$. We can easily show that
Eq. (\ref{I4}) holds for all anisotropy $k$. Such result is different from
the case considered in \cite{aharony}, where the amplitude ratio depends on $k$.
\end{references}
\end{document}